\documentclass[aps,superscriptaddress]{revtex4}

\usepackage{color}
\usepackage{graphicx}
\usepackage{amssymb}
\usepackage[T1]{fontenc}
\usepackage[latin9]{inputenc}



\begin{document}

\title{Spin entanglement, decoherence and Bohm's EPR paradox}

\author{E. G. Cavalcanti}

\affiliation{Centre for Quantum Dynamics, Griffith University, Brisbane QLD 4111, Australia}

\author{P. D. Drummond}

\affiliation{ARC Centre of Excellence for Quantum-Atom Optics, Swinburne University of Technology, Melbourne VIC 3122, Australia}

\author{H. A. Bachor}

\affiliation{ARC Centre of Excellence for Quantum-Atom Optics, Building 38, The Australian National University, Canberra ACT 0200, Australia}

\author{M. D. Reid}

\affiliation{ARC Centre of Excellence for Quantum-Atom Optics, Swinburne University of Technology, Melbourne VIC 3122, Australia}

\begin{abstract}
We obtain criteria for entanglement and the EPR paradox for spin-entangled
particles and analyse the effects of decoherence caused by absorption and state purity
errors. For a two qubit photonic state, entanglement can occur for
all transmission efficiencies. In this case, the state preparation purity
must be above a threshold value. However, Bohm's spin EPR paradox
can be achieved only above a critical level of loss. 
We calculate a required efficiency of 58\%, which appears achievable with current quantum optical
technologies. For a macroscopicnumber of particles prepared in a correlated state, spin entanglement and
the EPR paradox can be demonstrated using our criteria for efficiencies $\eta>1/3$ and
$\eta>2/3$ respectively. This indicates a surprising insensitivity
to loss decoherence, in a macroscopic system of ultra-cold atoms or
photons.
\end{abstract}
\maketitle

%\ocis{(270.5585) Quantum information and processing;
%(270.6570) Squeezed states; (020.1475) Bose-Einstein condensates. } % REPLACE WITH CORRECT OCIS %CODES FOR YOUR ARTICLE

\section{Introduction}

In the development of quantum information science, entanglement is
central. It is at the heart of the Einstein-Podolsky-Rosen (EPR) paradox\cite{einstein}
and Bell's theorem\cite{Bell}, which draw a clear delineation between
local realistic and quantum theories. Entanglement is also considered
a vital resource for future quantum technologies, both for photonic
systems and for applications involving ultra-cold quantum gases. 

Crucial to the generation and detection\cite{Peres,Horodecki,wit2,wit3}
of entanglement is decoherence\cite{ebscience}, which is the degradation
of a pure state into a mixed state due to coupling with the environment.
Decoherence can degrade or even destroy entanglement. Sensitivity
to decoherence, particularly for large systems, is thought to explain
the transition from the quantum to classical regime\cite{zureck}.
However, decoherence can be caused by many different physical mechanisms,
including particle loss, phase errors, and mixing with uncorrelated
particles.

This leads to the following fundamental question. When is entanglement
and EPR correlation preserved between two $N$ particle systems, if
each is independently decohered? Yu and Eberly\cite{yueberly1, yueberly2} studied
the issue for $N=1$, for a certain type of decoherence, and showed
that entanglement can be destroyed at a finite time. This is termed
{}``entanglement sudden death'' (ESD). The existence of ESD -- which
is experimentally verified\cite{alexp} -- has far reaching implications
for quantum information, since error correcting protocols may restore
a degraded but nonzero entanglement\cite{errocorr1,errocorr2,errocorr3}. However, these
questions have not been investigated in detail for EPR correlations,
which are more sensitive to decoherence than entanglement per se.

In this communication, we obtain quantitative criteria
applicable to Bohm's original  two-particle spin
realization\cite{bohm} of the EPR state, and generalize these to 2N-particle states
which display spin entanglement.
We investigate both EPR correlations and entanglement for these correlated multi-particle
states.  Two distinct types of decoherence are investigated. We show that entanglement
can resist decoherence even for the case of large $N$. To understand
this feature, we distinguish between \emph{noise} and \emph{loss} decoherence. With
 \emph{noise} decoherence,
the wrong information ({}``up'' instead of {}``down'') is
given, while \emph{loss} decoherence  causes an absence of information
and broadening\cite{Lars} (e.g. the changing of a qubit into a qutrit in the lossy situation
of section 3) of the measured Hilbert space. 

The decoherence
causing entanglement sudden death is essentially \emph{noise}. It leads to density matrices that are mixtures
with random states, with state purity $R=Tr\left[\rho^{2}\right]<1$.
In the Werner state\cite{werner,franswerner}, for example, the entangled two qubit
Bell state is mixed with a random state, and complete disentanglement
occurs once $R\leq1/3$\cite{Peres}. 

\emph{Loss} decoherence gives a completely different result. In the
case of the photonic two qubit Bell state\cite{franswerner,efficiency,Efficiency2},
used extensively in many seminal experiments and applications, loss
arises from absorption of photons into the environment, and is the
major source of decoherence. This loss causes an absence of a {}``count'', the latter 
arising with probability $\eta$.

The key result of the present analysis is that loss in the case of
a Bell state only \emph{reduces} the entanglement, which remains detectable
at all $\eta>0$. We extend this to treat EPR correlations, which
are more sensitive to decoherence. Both results are more experimentally
accessible than the violation of Bell inequalities, which require
high efficiencies ($\eta>.83$)\cite{Garg1987,Lars}. We also treat systems with macroscopic particle number. Such cases
have significance both as fundamental tests of quantum measurement
theory, and as potentially important quantum technologies for measurements both with photons and with ultra-cold 
bosonic atoms in correlated states.

\begin{figure}
\includegraphics[width=1\columnwidth]{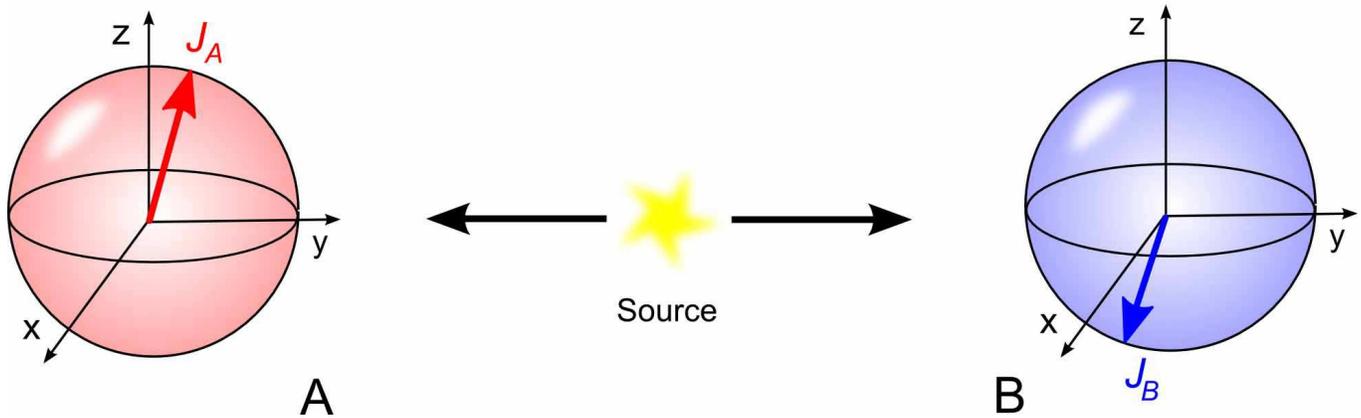}

\caption{Schematic diagram of Bohm's EPR experiment with correlated spins at
spatially-separated locations A and B.}

\end{figure}

\section{Bell state entanglement with losses}

We start with Bohm's gedanken-experiment  --- often called the Bell\cite{Bell} singlet state. This could involve
any particles having internal degrees of freedom, for example photons
or atoms. In the laboratory, there is noise in the form of randomly
polarised particles, and various forms of loss that cause only one
or zero particles to be detected, instead of two. We describe noise
using a Werner\cite{werner} state. This is composed of a singlet
Bell state (Fig. 1), 

\begin{equation}
\left|\Psi\right\rangle _{S}=\frac{1}{\sqrt{2}}\left(\left|1\right\rangle _{A}\left|-1\right\rangle _{B}-\left|-1\right\rangle _{A}\left|1\right\rangle _{B}\right)\,\label{eq:bellstate}\end{equation}
 with probability $p$, and a state with a particle of random spin
at each detector, with probability $1-p$. In the photonic case, $\left|\pm1\right\rangle _{A}$
indicates a photon with positive or negative helicity detected at
$A$, or more generally simply two distinct spin states of any quantum
field. The overall Werner state is then $\rho_{W}=p\rho_{S}+(1-p)I_{4}/4$.
Here $I_{4}$ indicates an identity operator on the two-particle subspace
of the four-mode Hilbert space. It includes all states with one particle
at $A$, and one at $B$, irrespective of their spin. The purity of
the Werner state is $R_{W}=Tr\left[\rho_{W}^{2}\right]=\left(3p^{2}+1\right)/4$.

We account for all loss that occurs prior to the measurement of the
{}``spin'' (polarization) of each particle\cite{detloss}, by defining
the overall efficiency as $\eta$. Thus at each detector, three outcomes
are possible: $+1$ (spin {}``up''); $-1$, (spin {}``down'');
and $0$, (no detection). The detection subspace corresponds to a
qutrit. To determine the density matrix, we derive the full matrix
based on a beam splitter model of loss\cite{ys}, in which the initial
state is represented as $\rho_{W}\rho_{vac}$. Here $\rho_{vac}=|0\rangle\langle0|$
is the multimode vacuum state for four field modes ($a_{\pm,vac}$
and $b_{\pm,vac}$) that collect lost photons. Thus, we assume the
standard quantum description of losses.

It is useful to adopt the Schwinger representation of the Werner state,
for which $|1\rangle\equiv|1,0\rangle_{A/B}$ and $|-1\rangle_{A/B}\equiv|0,1\rangle_{A/B}$
where $|i,j\rangle_{A}$ means $i$ and $j$ quanta in two distinguishable field modes
at $A$ that have spin labels $+1$ and $-1$, and for which $a_{\pm}^{\dagger}$
are the creation operators respectively. States at $B$ for modes
$b_{\pm}$ are defined similarly. We call the $A$ and $B$ measurements
Alice's and Bob's respectively.

The effect of the beam splitter model is to couple the field and vacuum
modes. After loss, the modes are transformed as $a_{\pm}\rightarrow\sqrt{\eta}a_{\pm}+\sqrt{1-\eta}a_{\pm,vac}$
and $b_{\pm}\rightarrow\sqrt{\eta}b_{\pm}+\sqrt{1-\eta}b_{\pm,vac}$.
We derive $\rho_{F}$, the matrix for the detected system obtained
by taking the trace over the lost photon modes. The $9$ basis states
are in three groups, categorised as $2$, $1$, $0$, by the number
of particles: $u_{1-4}=|\pm1\rangle_{A}|\pm1\rangle_{B}$; $u_{5,6}=|\pm1\rangle_{A}|0\rangle_{B}$,
$u_{7,8}=|0\rangle_{A}|\pm1\rangle_{B}$; and $u_{9}=|0\rangle_{A}|0\rangle_{B}$.
We find that:\begin{equation}
\rho_{F}=\left[\begin{array}{ccc}
\eta^{2}\rho_{W} & 0 & 0\\
0 & (\eta/2)(1-\eta)I_{4} & 0\\
0 & 0 & (1-\eta)^{2}\end{array}\right]\,\,.\end{equation}

Whether the decohered system $\rho_{F}$ is entangled is readily determined
using the PPT criterion\cite{Peres,Horodecki}, that if a partial
transpose of the density matrix has a negative eigenvalue, then the
state must be entangled. Calculation of the eigenvalues reveals entanglement
for all $\eta>0$ and $p>1/3$. A plot of the negativity (magnitude of the smallest negative eigenvalue of the density matrix) $\eta^{2}(3p-1)/4$ is shown in Fig. 2. This could be experimentally determined using quantum state
tomography.

\begin{figure}
\includegraphics[width=0.5\columnwidth,angle=270]{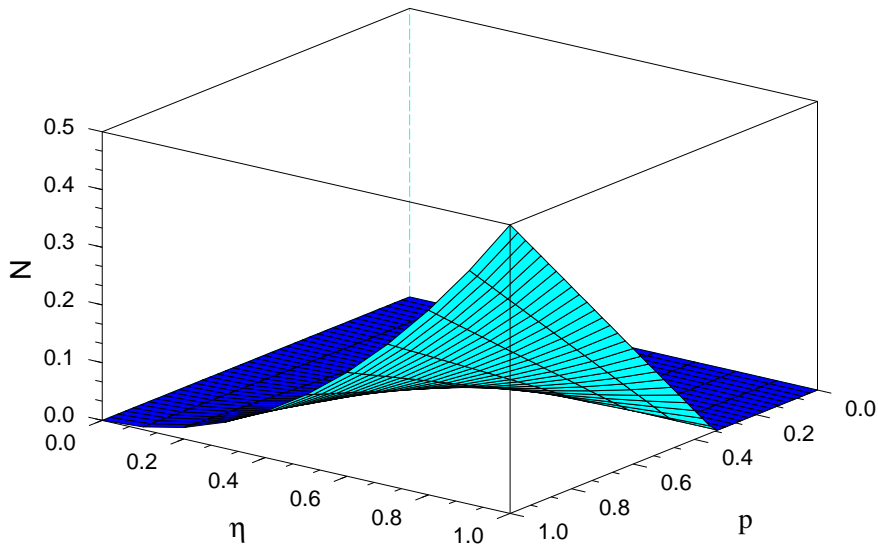}

\caption{Negativity of decohered Bell state to show entanglement sudden death
for \emph{noise} decoherence at $p<1/3$, but continuous suppression
of entanglement for \emph{loss} decoherence at all $\eta$. }

\end{figure}

We emphasize that here we proceed by taking a standard quantum theoretic
approach to loss. Within our quantum treatment, $\eta$ incorporates
the effect of all generation and detection losses. It is also possible
to take a black box approach, and consider a completely unknown cause of measurement
errors\cite{detloss}, which leads to a different efficiency threshold.

\section{Projected entanglement measures}

We now show that a more practical criterion to confirm entanglement
is obtained using operators which project onto the subspace in which
no photons are lost. Experimentally, this amounts to the procedure
of measuring the {}``qubit'' at each detector $A$ and $B$ only
where one obtains a count at each detector. This procedure is exploited
in many seminal experiments\cite{aspect} that infer the properties
of the photonic Bell state. We now formally validate this approach
when inferring entanglement, for the case where photons are lost prior
to measurement\cite{detloss}.

We use the Schwinger representation approach to define spin operators
in terms of the particle numbers detected at each location\cite{detloss}:
$J_{x}^{A}=(a_{+}^{\dagger}a_{-}+a_{-}^{\dagger}a_{+})/2$, $J_{y}^{A}=i(a_{-}^{\dagger}a_{+}-a_{+}^{\dagger}a_{-})/2$,
$J_{z}^{A}=(a_{+}^{\dagger}a_{+}-a_{-}^{\dagger}a_{-})/2$ . The total
observed particle number operator at Alice's location $A$ is $N^{A}=a_{+}^{\dagger}a_{+}+a_{-}^{\dagger}a_{-}$.
At Bob's location, $B$, $J_{x}^{B}$, $J_{y}^{B}$, $J_{z}^{B}$
and $N^{B}$ are defined in terms of $b_{\pm}$. Since $N^{A,B}$=0,1
these are also projectors $P^{A,B}$ on the subspace of particles
detected at $A,B$ respectively. Results for $J_{Z}$ at $A$ can
be $\pm1$, $0$, where $0$ is only achieved where there is no photon
counted. We now suppose Alice measures the local projection $s_{z}^{A}=J_{z}^{A}P^{A}$.
She thus defines her spin operators after projection onto the spin-$1/2$
subspace: $s_{x}^{A}=J_{x}^{A}P^{A}=(1/2)[|10\rangle\langle01|+|01\rangle\langle10|]_{A}$,
$s_{y}^{A}=J_{Y}^{A}P^{A}=(1/2i)[|10\rangle\langle01|-|01\rangle\langle10|]_{A}$,
$s_{z}^{A}=(1/2)[|10\rangle\langle10|-|01\rangle\langle01|]_{A}$
and $s_{A}^{2}=s_{X}^{2}+s_{Y}^{2}+s_{Z}^{2}=(3/4)N^{A}$. Similar
operators are defined for Bob.

Alice and Bob can also measure polarisations on a doubly projected
subspace, corresponding to coincident counts. Their measurements then
become nonlocal, given by the set $S_{\theta}^{A|B}=J_{\theta}^{A}P^{A}P^{B}=s_{\theta}^{A}P^{B}$
and $S_{A|B}^{2}=(3/4)P^{A}P^{B}$ where $\theta\in\{x,y,z\}$. We
have already defined the full $9\mathrm{x}9$ density matrix
$\rho$. The projected $4\mathrm{x}4$ one is $\rho_{proj}=P^{A}P^{B}\rho P^{A}P^{B}/Tr[\rho P^{A}P^{B}]$.
This projected density matrix, as one might expect intuitively, is
just the Werner density matrix that we started with.

We wish to prove that the projected measurements are sufficient to
prove entanglement over the full density matrix. This follows using
the fact that local projections or their products cannot induce entanglement\cite{Horodecki}.
Hence, a criterion sufficient for entanglement on the projected density
matrix is also sufficient over the complete density matrix. As an
example, we consider the entanglement criterion of Hofmann and Takeuchi\cite{hof},
and define a sum of spin measurements for Alice and Bob as $s_{\theta}=s_{\theta}^{A}+s_{\theta}^{B},$
where $\theta\in\{x,y,z\}.$ It is therefore sufficient to measure
$\Delta^{2}s_{x}+\Delta^{2}s_{y}+\Delta^{2}s_{z}<1$ over the two-photon
subspace to confirm entanglement on the full space. This is always
possible\cite{hof} provided $p>1/3$. 

It is perhaps rather obvious that by losing some particles one can
still retain entanglement. After all, if one has several copies of
an entangled state, but some of them are lost, the resulting ensemble
should still be entangled. However, it might also seem {}``obvious"{}
that if one has an ensemble of entangled states and some of them are
replaced by mixed states, the ensemble should still retain some entanglement.
In reality this is actually false. The interest of the apparently
{}``obvious'' (and true) result about entanglement with loss must
be contrasted with the falsity of the also apparently {}``obvious''
result about noise.

Our conclusion is therefore that as far as entanglement is concerned,
the effect of particle losses on the Bell state can be ignored, if
one simply makes measurements conditioned on observing two-particle
coincidences. This is, of course, a key difference between the two kinds of decoherence: the effect of loss can be filtered out by post-selecting the subset of measurements in which all expected detections occur, whereas this cannot be done for noise. In other words, \emph{loss }decoherence has no effect
on entanglement --- there is no ESD here --- while \emph{noise} decoherence
has a stronger effect, causing a decoherence threshold which is equivalent
to the ESD phenomenon. In both cases, the total density matrix prior
to measurement is in a mixed state caused by the decoherence.

\section{Spin EPR paradox}

Next, we turn to the EPR paradox. This is much more challenging experimentally.
The paradox shows that local realism (LR) is inconsistent with the
completeness of quantum mechanics, which is a stronger result than
entanglement. As a first requirement, since EPR's no {}``action-at-a-distance''
is crucial to the local realism part of the EPR argument, one must
have causal separation between Alice's and Bob's measurements. Thus,
for EPR, we \emph{must rule out the nonlocal procedure of projections}
onto the two-photon subspace\emph{.} Alice's and Bob's measurements
must be \emph{local}. Second, the measurements at Alice's location
must allow a local state to be inferred at Bob's location --- assuming
LR. If this inferred state has a lower uncertainty than allowed by
quantum mechanics, the EPR paradox is obtained. Thus LR is false (local
prediction does not imply a local element of reality) \emph{or} quantum
mechanics is incomplete (it fails to fully describe the inferred state,
since this violates the uncertainty principle). This logic is central
to the EPR argument applied to real experiments.

It is important here to recognise the difference between the EPR and
Bell arguments. Following Einstein, we explicitly assume that quantum
theory correctly describes our measurements. The EPR logic does not
require any alternative theory to quantum theory. It simply deals
with the question of whether the completeness of quantum mechanics
is compatible with local realism. Therefore, there is no need here to consider
Bell's investigation into local hidden variable theories, which may have
an arbitrary treatment of loss. This means that we can use a standard
quantum treatment of loss.

Our route to a signature of an unambiguous EPR paradox\cite{eprr1,eprr2}
is via an inference argument together with the known quantum uncertainty
principle $\Delta^{2}J_{x}+\Delta^{2}J_{y}+\Delta^{2}J_{z}\geq\langle N\rangle/2$
for spins --- the same uncertainty principle used in the derivation
of the entanglement criterion of \cite{hof,entsig}, given as 
\begin{equation}
\Delta^{2}J_{x}+\Delta^{2}J_{y}+\Delta^{2}J_{z}<\langle N^{A}+N^{B}\rangle/2, \label{eq:entcond}
\end{equation}
where $J_{\theta}=J_{\theta}^{A}+J_{\theta}^{B}$. Our EPR criterion
simply requires that the inferred variance of Bob's measurements must
be less than that for any possible quantum state; that is:
\begin{equation}
\Delta_{inf}^{2}J_{x}^{B}+\Delta_{inf}^{2}J_{y}^{B}+\Delta_{inf}^{2}J_{z}^{B}<\langle N^{B}\rangle/2. \label{eq:eprcond}\end{equation}
 Measurement schemes for all quantities in (\ref{eq:eprcond}) have
been demonstrated in recent polarisation-squeezing experiments\cite{spinpolepr1,spinpolepr2}.
Here $\Delta_{inf}^{2}J_{x}^{B}=\sum_{J_{x}^{A}}P(J_{x}^{A})\Delta^{2}(J_{x}^{B}|J_{x}^{A})$
is the average, over $J_{x}^{A}$, of the conditional variances $\Delta^{2}(J_{x}^{B}|J_{x}^{A})$,
for a measurement $J_{x}^{B}$ given an outcome $J_{x}^{A}$. This
inferred uncertainty is the average error associated with the inferred
result for a remote measurement $J_{x}^{B}$, given measurement of
$J_{x}^{A}$. To prove the EPR criterion (\ref{eq:eprcond}), one
considers the conditional distributions as predictions for $B$ given
$A$\cite{eprr1,eprr2}. If LR holds, the predetermined prediction for $J_{x}^{B}$
means there is a corresponding localised state $\rho_{B}$ at $B$.
This is because if the systems are causally separated, according to
LR, the measurement at $B$ does not induce immediate change to $A$.
EPR called such predetermined states {}``elements of reality''.

In the case of Bohm's EPR paradox, the assumption of LR means that
elements of reality exist for each of the spins $J_{x}^{B}$, $J_{y}^{B}$,
$J_{z}^{B}.$ The variances associated with the prediction for each
of them are respectively, $\Delta_{inf}^{2}J_{x}^{B}$, $\Delta_{inf}^{2}J_{y}^{B}$,
$\Delta_{inf}^{2}J_{z}^{B}$. Where we satisfy (\ref{eq:eprcond}),
EPR's elements of reality defy the quantum uncertainty relation for
$B$. That is, it is impossible to represent Einstein's proposed element
of reality as a quantum state $\rho_{B}$. In this way the EPR paradox
is able to be experimentally demonstrated. This is an important conceptual
boundary, which demonstrates the inadequacy of the classical concept
of local realism in dealing with quantum states.

When loss is included, we find that it increases the uncertainties
associated with the inference of measurements at Bob's location. As
before, we take the Werner state and calculate the EPR inequality
with loss included. The RHS of (\ref{eq:eprcond}) is $\eta/2$ while
$\Delta_{inf}^{2}J_{x}^{B}=\Delta_{inf}^{2}J_{y}^{B}=\Delta_{inf}^{2}J_{z}^{B}=\eta(1-\eta^{2}p^{2})/4$.
The EPR criterion is then satisfied for $\eta p>1/\sqrt{3}$. This
implies that, unlike the entanglement case, both loss and noise have
a similar effect on the EPR paradox. The reason is simply that the
EPR paradox is related to causality. Nonlocal projections cannot be
used, as in the entanglement case, to obtain a smaller ensemble for
conditional measurement. This is also the same reason why one cannot
use the term EPR paradox or Bell inequality unless there is a clear
causal separation between the measurement events.

Although the required efficiency is greater than in any reported Bell
state measurement to date, it is within reach of current
photo-detectors. It would be an interesting challenge to demonstrate
the EPR paradox for spatially separated, correlated particles. This
would resolve Furry's question\cite{furry} about the possibility of entanglement decay
for separated massive particles, which was an early proposal to resolve
the EPR paradox.

\section{Macroscopic EPR entanglement}

Finally, we consider entanglement and EPR for macroscopic states with
\emph{more} than one particle per mode. This implies that we now consider
only bosonic fields, like photons or ultra-cold BEC experiments. These
correlated states would give a much more powerful test of quantum
measurement theory, testing features of quantum reality in domains
that become meso- or macroscopic. In this domain, a number of alternatives
to quantum mechanics have been suggested, where quantum superpositions
are prevented from forming via novel mechanisms such as couplings
to gravitational fields\cite{Gravity}. If gravitational effects are
involved, it seems clear that one must test the relevant quantum predictions
for massive particles, in order to allow for a strong enough gravitational
coupling to occur.

To test for quantum effects in such macroscopic cases, we first consider
the way in which the relevant states would be generated in practice.
We consider a macroscopic version of the Bell state (\ref{eq:bellstate}),
using the Schwinger representation \textcolor{black}{\begin{eqnarray}
{\normalcolor {\color{red}{\normalcolor |\psi_{N}\rangle}}} & = & \frac{1}{N!\sqrt{N+1}}(a_{+}^{\dagger}b_{-}^{\dagger}-a_{-}^{\dagger}b_{+}^{\dagger})^{N}|0\rangle.\label{eq:drum}\end{eqnarray}
} 

We can generate the states of Eq. (\ref{eq:drum})
using two parametric amplifiers\cite{spinpolepr1,spinpolepr2} as modeled by the interaction Hamiltonian
\textcolor{black}{\begin{equation}
H=i\hbar\kappa(a_{+}^{\dagger}b_{-}^{\dagger}-a_{-}^{\dagger}b_{+}^{\dagger})-i\hbar\kappa(a_{+}b_{-}-a_{-}b_{+}).\label{eq:parampham}\end{equation}
 With an initial vacuum state, the solution after a time $t$ is a
superposition of the $|\psi_{N}\rangle$.} In the regime of large
$\langle N^{B}\rangle$, higher photodiode detection efficiencies
($\eta\approx0.9$) can be achieved, although a precise photon count,
which would enable a test of Bell's inequality\cite{bellspin1,bellspin2,bellspin3}, is
difficult. The solutions are readily obtained to give 
\begin{eqnarray}
a_{\pm}&=&a_{\pm}(0)\cosh(r)\pm b_{\mp}^{\dagger}(0)\sinh(r)\\
b_{\pm}&=&b_{\pm}(0)\cosh(r)\mp a_{\mp}^{\dagger}(0)\sinh(r),
\end{eqnarray} 
where
$a_{\pm}(0)$ represent vacuum initial states and $r=|\kappa|t$. The effect of loss
is analysed using a standard beam splitter model\cite{ys} which adds
vacuum terms so that final outputs after loss become $a_{\pm}^{L}=\sqrt{\eta}a_{\pm}+\sqrt{1-\eta}a_{\pm,0}$
and $b_{\pm}^{L}=\sqrt{\eta}b_{\pm}+\sqrt{1-\eta}b_{\pm,0}$ where
the $a_{\pm,0}$ and $b_{\pm,0}$ represent independent vacuum inputs.
With this we get 
\begin{eqnarray}
\langle(J_{Z}^{A})^{2}\rangle&=&(1/2)\eta\sinh^{2}(r)(1+\eta\sinh^{2}(r))\\
\langle J_{Z}^{A}J_{Z}^{B}\rangle&=&-(1/2)\eta^{2}\cosh^{2}(r)\sinh^{2}(r),
\end{eqnarray} 
which gives a final result for the spin uncertainties of:
\begin{equation}
\Delta^{2}J_{x}=\Delta^{2}J_{y}=\Delta^{2}J_{z}=\eta(1-\eta)\sinh^{2}(r)\,,
\end{equation}
and $\langle N\rangle=4\eta\sinh^{2}(r)$. For all $N$, efficiencies $\eta>1/3$ are enough
to demonstrate entanglement.

\begin{figure}
\includegraphics[width=10cm]{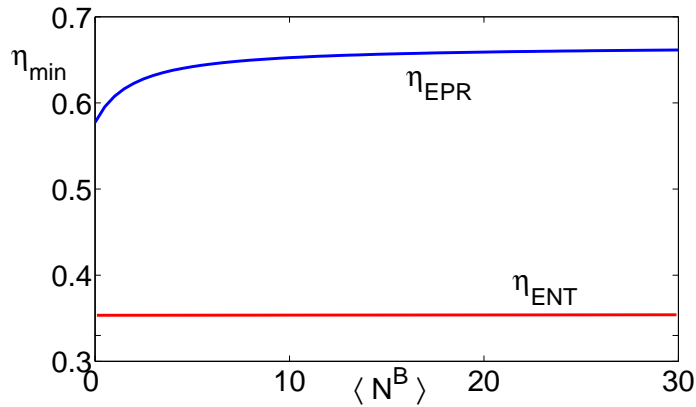}~

\caption{\textcolor{black}{(a) Threshold detection efficiency $\eta_{min}$
required (by (\ref{eq:parampham})) to confirm entanglement and the
spin EPR paradox via (\ref{eq:entcond}) and (\ref{eq:eprcond}),
for a given mean photon number $\langle N^{B}\rangle$.}}

\end{figure}

A similar result is obtained for the spin EPR correlations. Here the
minimum efficiency required to satisfy (\ref{eq:eprcond}) approaches
$\eta=2/3$ for infinite $\langle N^{B}\rangle$. We calculate the
conditional variances for this Gaussian system using \textcolor{black}{a
linear regression approach\cite{eprr1,eprr2}, where the estimate for the
result of the remote measurement $J_{x}^{B}$ is simply $J_{\theta,est}^{B}=gJ_{\theta}^{A}$,
so that the average inference variance is $\Delta_{inf}^{2}J_{\theta}^{B}=\langle(J_{\theta}^{B}-gJ_{\theta}^{A})^{2}\rangle$.
We calculate the linear inference variance for (\ref{eq:parampham})
by selecting $g=-\langle J_{\theta}^{B}J_{\theta}^{A}\rangle/\langle J_{\theta}^{A}J_{\theta}^{A}\rangle$
to minimise $\Delta_{inf}^{2}J_{\theta}^{B}$: 
\begin{eqnarray}
\Delta_{inf}^{2}J_{\theta}^{B} & = & \langle(J_{\theta}^{B})^{2}\rangle-\langle J_{\theta}^{B}J_{\theta}^{A}\rangle^{2}/\langle(J_{\theta}^{A})^{2}\rangle\nonumber \\
 & = & \frac{\eta sinh^{2}r(1-\eta^{2}+2\eta(1-\eta)sinh^{2}r)}{2(1+\eta sinh^{2}r)}\label{eq:infer}
 \end{eqnarray}
and $\langle N^{B}\rangle=2\eta sinh^{2}r$. Figure 3 plots the threshold
efficiency $\eta$ for satisfaction of (\ref{eq:eprcond}), to indicate
a test of macroscopic EPR for large $\langle N^{B}\rangle$ and $\eta>2/3$.}

Note that the calculations in this section refer to {\em loss} decoherence. A calculation including a model of the effect of {\em noise} in the entanglement of this many-particle system would of course be important before experimental realisation of this proposal. As for the two-qubit case, we expect that some finite amount of noise will lead to the elimination of entanglement as well as the EPR paradox, with the precise value depending on the type of noise affecting the system.

\section{Conclusion}

We have shown that it is possible to demonstrate two qubit entanglement
for any value of loss, although ESD occurs when there is noise decoherence.
Demonstrating Bohm's two qubit spin EPR paradox\cite{bohm} is more
difficult. With our criterion, this is only possible above a critical detection efficiency
 $\eta>1/\sqrt{3}$. This is still more accessible than a loophole-free demonstration of Bell nonlocality, which has an even higher efficiency threshold.

The significance of progressively testing for stronger forms of nonlocality,
from entanglement to the EPR paradox through to Bell's theorem, has
been outlined recently by Wiseman {\em et al.}\cite{hw}, who report a cut-off
of the EPR paradox for Werner states, at $p\leq0.5$.

We then progress to examine the resilience of the nonlocality of \emph{macroscopic}
systems to decoherence. We report that the entanglement and Bohm's
spin EPR paradox are preserved for $\eta>1/3$ and $\eta>2/3$ respectively,
even for higher qubit systems with arbitrarily large $N$. This is
a surprising result that contradicts the popular view that sensitivity
of entanglement to decoherence increases with the {}``largeness''
of bodies entangled\cite{ebscience,ys}.

Our prediction that entanglement between macroscopic (arbitrary $\langle N^{B}\rangle$)
systems is preserved up to a large and fixed loss appears to counter
previous results regarding macroscopic decoherence \cite{zureck,ys,yueberly1,yueberly2}.
Yet, the result is consistent with recent predictions for decoherence
based on mixing with noisy states\cite{bellnoise}, and reports of
experimental measurement of entanglement between large, lossy systems\cite{bow,spinpolepr1,spinpolepr2,atomicens}.
The prediction can be tested with either photonic or massive atomic
systems\cite{Gravity,massive2}, leading both to new understandings
and tests of quantum mechanics, and the possibility of novel quantum
technologies.

In current optical experiments the best quantum efficiency achievable is about 75\%, which is typically a combination of losses in the apparatus (85\%)\cite{Generator}, mode matching (95\%) and efficiency of the detectors (95\%)\cite{Photodetector}. Further refinements, like those employed in the best optical squeezing experiments \cite{NinedB, TendB},  will bring the total efficiency closer to 90\%. This is well above our calculated benchmark efficiency of 58\% for an EPR test, provided other technical noise sources can be suppressed sufficiently well. In summary, an unambiguous experimental test of this EPR criterion is not impossible.

\section*{Acknowledgments}
 This work was funded by the Australian Research Council
Center of Excellence for Quantum-Atom Optics, a Griffith University Postdoctoral Research Fellowship, an ARC Postdoctoral Research Fellowship and an ARC Professorial Fellowship .

\end{document}